\documentstyle[preprint,prb,aps] {revtex}
\begin{document}
\everymath{\rm}
\begin{center}
{\Large {\bf On the Josephson coupling between a disk of one superconductor
and a surrounding superconducting film  of different symmetry}} \\
\bigskip
A. J. Millis\\
\bigskip
AT\&T Bell Laboratories\\
600 Mountain Ave\\
Murray Hill, NJ 07974\\
\bigskip
{\it ABSTRACT}
\end{center}
A cylindrical Josephson junction with a spatially dependent Josephson coupling
which averages to zero is studied in order to model the physics of
a disk of d-wave superconductor embedded in a
superconducting film of a different symmetry.
It is found that the system always introduces
Josepshon vortices in order to gain
energy at the junction.
The critical current is calculated.
It is argued that a recent experiment claimed to
provide evidence for s-wave superconductivity in $YBa_2Cu_3O_7$
may also be consistent with d-wave superconductivity.
\newpage

In this note I analyze a theoretical model of an experimental
system recently proposed as a test for the presence of
d-wave superconductivity in the cuprate high-$T_c$
superconductors [1].
The proposed experimental system is shown schematically
in figure 1.
[The inner region is drawn  as a circle, the
region used in ref. 1 was a hexagon.]
The  inner region, labelled A, is a c-axis-oriented
film of high-$T_c$ $CuO_2$ superconductor with the Cu-O
bonds running vertically and horizontally as shown.
In the devices actually made [1] the outer region,
B, is another c-axis film of high-$T_c$
superconductor, with the Cu-O bonds rotated by $45^{\circ}$
with respect to region A, as shown.
I will also consider the case in which  region B is
an s-wave superconductor.
In either case the Josephson coupling alternates in sign as
one proceeds around the perimeter of region $A$, and averages to zero.
The experiment is to measure the critical current of the A-B
interface.
A nonzero critical current was argued [1] to be evidence
against d-wave pairing because
if one chooses a constant
phase $\phi_A$ in region
A and a constant phase $\phi_B$ in region B, then
the Josephson energy of the interface vanishes and so there
can be no supercurrent across it.
This argument is incorrect;
the difficulty is the assumption of constant,
spatially independent phases $\phi_A$ and $\phi_B$.
I shall show that for the two dimensional geometry of interest, the system
introduces Josephson vortices to gain  energy at the boundary
at the expense of introducing bulk supercurrents.  The physics is similar
to that of inductive effects in a SQUID, where one balances the Josephson
energy against the magnetic field energy.  The dimensionality is crucial
to the argument; the results do not apply to the essentially one
dimensional
"corner junction" SQUID experiments recently argued [2] to provide evidence
of d-wave superconductivity in $YBa_2Cu_3O_{7-\delta}$.

In the remainder of this communication I analyze the interplay between
bulk currents and the Josephson energy more quantitatively,
by minimizing the appropriate free energy.
The free energy has three contributions:
the Josephson energy, the supercurrent energy, and the energy in the
magnetic field generated by the supercurrent flow.
I first study the simplest case, that of a very
thin film for which the penetration depth may be taken to
be infinite and all magnetic effects are negligible.
Then I consider the general case, and finally discuss the
relation to experiments.

Assume first that the only contributions
to the energy are the bulk supercurrents (i.e. the kinetic inductance)
and the Josephson
energy of the A-B interface.
The energy of a bulk supercurrent involves
a superfluid stiffness $\rho$ which I relate to a bulk critical
current $J_c$ by the argument
that at the critical current the phase changes by an
amount of order 1 in a distance of order the coherence length,
$\xi_0$, i.e. $J_c \xi_0=\rho$.
The energy, $E$, is then
$$
E  =  \int_0^a rdrd\theta \frac{1}{2}
J_c^A \xi_0^A (\bigtriangledown \phi^A)^2+
\int_a^{\infty} rdrd\theta
\frac{1}{2} J_c^B \xi_0^B  (\bigtriangledown \phi^B)^2
$$
$$
+ I_J a \int_0^{2\pi}
d\theta g_M(\theta ) \cos
[\phi^A (r=a,\theta ) -
\phi^B (r=a,\theta )  ]
\eqno{(1)}
$$

Here $\phi^{A,B}$ are the
superconducting phases in
regions $A$ and $B$ respectively, $I_J$ is the maximum
Josephson current of an infinitesimal region of the
interface, and $g_M(\theta)$ describes the angular dependence of the
Josephson coupling.
In the case where $A$ is d-wave and $B$ is s-wave, we
must on general grounds have
$\int_0^{2\pi} d\theta g_M ( \theta ) = 0$ and
$g_M(\theta + \pi/2) = -g_M(\theta )$.  I call this case M=2.
In the case considered in ref. (1) we must have
$\int_0^{2\pi} d\theta g_M (\theta) = 0$ and
$g_M(\theta + \pi/4)=-g_M(\theta)$; I call this M=4.
For explicit calculations I take the simplest forms
consistent with these considerations[3,4], namely
$g_{2}(\theta) =  \sin 2\theta \; \; $ or
$g_{4}(\theta ) = \sin 4\theta \; \; $. I indicate the modifications
arising for the alternative case of a piecewise constant
$g _M(\theta)$.

The theoretical problem is to minimize the
energy in (1) subject to the constraints of local current
conservation at each point of the A-B interface and fixed total radial
current.
Local current conservation implies ($\hat{n}$ is the outward
normal to the A-B interface at angle $\theta$)
$$
J_c^{A}
\xi_0^A
\hat{n} \cdot \vec{\bigtriangledown} \phi^A   =
J_c^B \xi_0^B  \hat{n} \cdot \vec{\bigtriangledown} \phi^B \\
\eqno{(2a)}
$$
$$
= I_J g(\theta) \sin [\phi^A (a,\theta)-\phi^B
(a,\theta) ]
\eqno{(2b)}
$$
In terms of the mean radial current density at the A-B
interface, $I_0$, the constraint on the total radial current is
$$
I_0 = \frac{I_J}{(2\pi)}
\int_0^{2\pi} d\theta
g(\theta) \sin
[\phi^A (a,\theta) -\phi^B
(a,\theta)]
\eqno{(3)}
$$
To perform the minimization, note that
Equation 1 implies that $\phi^{A,B}$ satisfy
$\bigtriangledown^2 \phi = 0$.
Assume that the current $2\pi I_0a$ is introduced
symmetrically
in the center of region A.
The problem then has circular symmetry; the
general solution to the Laplace
equation may be written down, the solutions
in regions A and B may be related using 2a, and the
result substituted into (1).

The result is
$$
\frac{E}{\bar{J}_c\xi_0}
= \frac{1}{2}
\sum_n  n (c_n^2+d_n^2)
\eqno {(4)}
$$
$$
\; \; \; \; - \bar{I}
\int_0^{2\pi}
\frac{d\theta}{2 \pi} g_M (\theta) \cos [ \phi ( \theta ) ]
$$
where
$$
\phi ( \theta ) =
\phi^A (a,\theta) -
\phi^B (a,\theta) =
\phi^0 + \sum_n
[c_n \sin n\theta + d_n \cos n\theta ]
\eqno {(5)}
$$
and $\bar{J} = J_c^A J_c^B / (J_c^A+J_c^B)$
and $\bar{I} = \pi I_J a/ \bar{J} \xi_0$.

Minimizing (5) with respect to the $c_n$ and $d_n$
at fixed $\phi_0$ ensures that all of the
$n > 0$ Fourier components of the current conservation
equation 2b are satisfied.
The mean current $I_0$ (eq. 3) is then fixed by $\phi_0$,
which is the angular average of the phase drop across
the A-B interface.

{}From (4) it is clear that the important parameter is
$\bar{I}$; for $\bar{I} \ll 1$ the Josephson
energy is weak while for $\bar{I} \gg 1$ it is strong.
Further, $\bar{I}$ increases linearly with the perimeter
of the disk, because in two dimensions the energy of bulk
superflow $\int d^2r ( \bigtriangledown \phi )^2$ is scale invariant.
Consider first $\bar{I} \ll 1$.
Clearly the free energy may be expanded in powers of the $c_n$
and $d_n$.  Expanding to leading nontrivial order and minimizing gives
$$
I_0 =
\frac{I_J\bar{I} \sin 2 \phi_0}{2M}
\eqno {(6a)}
$$
and
$$
E=- \frac{\bar{I}^2}{4M}
\sin^2 \phi_0
\eqno {(6b)}
$$
Observe that $E$ is minimized at a phase
difference of $\pi / 2$.
The bulk supercurrents  at
$\phi_0= \pi /2$ are shown in fig. 2,
for $M=2$.
One sees that the solution corresponds to
2M vortices, with alternating positive and negative
circulation, centered at the points where
$g_M(\theta)$ vanishes.
Because $\bar{I} \ll 1$, the ``core''
of the vortex is much larger than the vortex spacing, $a$.
If a mean radial current is imposed,
$\phi_0 \neq \pi /2$ and the centers of the vortices shift along the
perimeter of the circle in such a way that the sum
of the imposed radial current and the circulating
current about each vortex vanishes at the points where
$g_M (\theta) = 0$.
The motion of the centers of the vortices is shown in
the inset to fig. 2.

In  the case of a piecewise
constant
$g_M (\theta)$, in which
$g_M(\theta = 1$ $(0 < \theta < \pi / M )$,
$g_M (\theta ) = -1$
($\pi / M < \theta < 2 \pi / M$) etc,
one finds
that the energy is still proportional to
$\sin^2 \phi_0$ but with a slightly different coefficient,
and the qualitative structure of the
vortices is not changed.

Note that the energy, eq. 6b, is periodic in $\phi_0$
with period $\pi$, unlike a conventional junction which would
have a $2 \pi$ periodicity.
A $\pi$-periodicity is unambiguous evidence of
unconventional superconductivity and may in principal be measured
in a SQUID experiment.
Now the quantity $\phi_0$ is the angular
average of the phase drop across the A-B
interface, but
from the solution of the Laplace equation that led to eq. (4)
I also find that at a distance $d \ll a$ from the
center of the disk, the phase has a negligible
angular dependence and is
$$
\phi (d) = \phi_0 -
\frac{\bar{I}}{\pi}
\; \; \ln \; a/d
\eqno {(7)}
$$
Thus for $\bar{I} \ll \pi/ln(a/d) \sim 1$  the phase in the
center of the disk
is essentially $\phi_0$ and so an experiment in which
this phase is controlled would measure a $\pi$ periodicity.

Now consider the opposite limit, $\bar{I} \gg 1$.
Specialize first to the case $\phi_0 = \pi /2$; then
it follows from the minimization equations that all the
$d_n=0$.
The Josephson term in eq. 4 is then minimized by a piecewise
constant $\phi ( \theta)$, equal to $\pm \pi/2$
according as $g ( \theta )$ is positive or negative.
Now a piecewise constant $\phi (\theta )$ implies
$c_n \sim 1/n$ and therefore a logarithmically divergent
$\sum_n nc_n^2$.
This divergence comes from the familiar $1/r$ behavior of
supercurrents outside the core of a vortex.
To cut off the logarithm one must convert the steps
in $\phi ( \theta )$ to smooth crossovers extending over a
scale $\epsilon$ which gives the
core size of the vortex.
For  $g_M(\theta) = sin(M\theta)$ the cost in Josephson
energy is $\sim \bar{I} \epsilon^2$, while the superfluid energy is
$\sim \ln 1/ \epsilon$.
Optimizing gives $\epsilon \sim \bar{I}^{-1/2}$ and
$E = -\bar{I} + O \ln \bar{I}$.
Thus at $\phi_0 = \pi /2$ and $\bar{I} \gg 1$ the system
gains essentially the maximum Josephson energy and the
vortices depicted in fig. 2 are pinched at the A-B interface
into angular regions of size $\bar{I}^{-1/2}$.
For a piecewise constant $g$,
the core size would be $\epsilon \sim \bar{I} ^{-1}$
and so the coefficient of the ln would change by a factor of two.

To treat general values of $\phi_0$ at $\bar{I} \gg 1$ one must
write $\phi(\theta)$ as a sum of two functions,
one constant in the interval
$0<\theta<\pi/M$ and the other piecewise constant and averaging
to zero (these represent the sin and cosine fourier components defined
in eq 5).  The details will be given elsewhere; the results is that
 up to terms of order
$(\ln ( 1 /\bar{I} )/ \bar{I} )^{2/3}$
the energy is independent of the imposed phase $\phi_0$.

This is related to the result of eq. (7),
namely that for $\bar{I} \gg 1$, the phase
for any current of order $I_J$ winds through may $2\pi$
revolutions between the center of the disk and the perimeter,
so the average phase difference $\phi_0$ is not an
easily controlled quantity.  To determine the critical
current one must
the energy subject to a constraint on the total current
(which is no longer
fixed by $\phi_0$; one finds that the critical current is  $2I_J/\pi$.

Note that in the $\bar{I} \gg 1$ case, unlike the
$\bar{I} \ll 1$ case, a solution with
$E < 0$ exists at $\phi_0 = 0$.
Therefore, as $\bar{I}$ is increased beyond some critical value,
$\bar{I}_c$, presumably of order 1 the $E < 0$ solution must
appear.
The transition region $\bar{I} \sim 1$ requires numerical
study, which has not
been undertaken.
However,  by expanding eq. (5) to quadratic order is the $c$'s
and $d$'s I have determined that the solution
$c_n = d_n = 0$ at $\phi_0 = 0$ becomes linearly
unstable at $\bar{I} = \bar{I}_c \cong 1.53$.  Note also that at large
$\bar{I}$ the $\pi$-periodicity which still exists in principle is of
no practical relevance because if one starts the system in the lowest
energy state at some initial phase $\phi_0$
and then adiabatically increases
the phase to $\phi_0+\pi$, the system will end up in a non-optimal
state (higher than the optimum by a relative energy of
order $ln(\bar{I})^{2/3}/\bar{I}$; to get to the optimal
state would require
overcoming an energy barrier of order $I_Ja$, which will be much
larger than $k_bT$ for a large device.

I now extend the treatment of to include the effects of a finite
London penetration depth, $\lambda$, and of the energy
stored in the magnetic field.
One expects these effects to be important because currents
in vortices decay exponentially on scales larger
than the effective penetration depth $\lambda_{eff}$
(which for a film of thickness $d < \lambda$ may be
greater than the microscopic penetration depth $\lambda$),
and because the magnetic field extends also in the third dimension.
The most interesting case is
$\lambda_{eff} \ll a$; the
currents in the vortices will then be confined to the region
close to the A-B interface,
so the precise geometry will not be important.
It is convenient to study the linear
interface shown in fig. 3.
Here regions A and B are sheets of thickness
d lying in the x-y plane and occupying the
half planes $x > 0$ and $x < 0$ respectively.
I assume the Josephson coupling
$I_J (y)$ along the A-B interface averages to zero,
is periodic with period 2a, and satisfies
$I_J (y+a) = -I_J(y)$.
To simplify the analysis I also assume that regions A
and B have the same material parameters.

I proceed as before, by first
determining the general form of the bulk
currents and then forcing this form to be
consistent with the Josephson relation across the
interface.
In the presence of a magnetic field $\vec{h}$ one
introduces a vector potential $\vec{A}$ given by
$\vec{\bigtriangledown} \times \vec{A} = \vec{h}$ and one
must use the gauge invariant supercurrent
$\vec{J} = \vec{\bigtriangledown} \phi -
\frac{2\pi\vec{A}}{\Phi_0}$
instead of $\vec{\bigtriangledown} \phi$.
It is also convenient to introduce a scaled field
$\vec{H}$ via
$\vec{H} = 2 \pi \vec{h} / \Phi_0$
Here $\Phi_0 = hc /2e$ is the superconducting flux
quantum.
In the volume occupied by the superconductor
the equations governing the superflow and field are:
$\vec{\bigtriangledown} \times \vec{J} = \vec{H}$ and
$\vec{\bigtriangledown} \times \vec{H} = \vec{J} / \lambda^2$.
Outside the film,
$\vec{J} = 0$ and $\vec{\bigtriangledown} \times \vec{H} = 0$
To simplify the analysis I assume that currents flow only in the x-y
plane.
This is equivalent to assuming that $d \gg \lambda$
or $d \ll \lambda$.
I give the analysis for the $d \gg \lambda$ case;
the $d \ll \lambda$ yields essentially identical results, but with
$\lambda$ replaced by
$\lambda_{eff} = \lambda^2 / d$.

For $d \gg \lambda$ the solution in the film is only weakly
perturbed by the spreading of the field in the region above
and below the film.
I therefore eliminate $\vec{H}$ from the in-film
equations, solve for $\vec{J}$,
compute $\vec{H}$ in the plane of the film, and use this
as a boundary condition for the equation for $\vec{H}$ in the
region outside the film.
The components of the current $\vec{J}$ are:
$$
J^x (x,y) =
\frac{\pi}{2a}
\sum_p p
\frac{b_p exp - \sqrt{1+b_p^2} |x| / \lambda}{\sqrt{1+b_p^2}}
$$
$$
\cdot [ c_p \; \sin
\frac{p \pi y}{a} +
d_p \cos \frac{p \pi y}{a} ]
\eqno (8a)
$$
$$
J^y(x,y) = \frac{\pi}{2a}
sgn (x) \sum_p exp -
\sqrt{1+b_p^2} |x| / \lambda
$$
$$
\cdot \left[ d_p \; \sin
\frac{p \pi y}{a} - c_p \cos \frac{\pi py}{a} \right]
\eqno (8b)
$$
where
$b_p = \pi p \lambda / a$.

To implement the Josephson boundary condition one must compute the
properly gauge invariant phase drop across the interface
$\Delta \phi_{GI}$.
Now $\Delta \phi_{GI}$ has a contribution $A^x (\Delta x)$ from
the change in the normal component of the vector potential
across the interface.   For a very thin junction,
$\Delta x \rightarrow 0$ and this contribution may be
obtained from the condition that the flux enclosed within the
contour shown in fig. 3 is negligible.  (The usual contour
extending to $|x| =\infty$ is not
convenient because of the vortex currents). I find $\Delta \phi_{GI} (y_2) -
\Delta \phi_{GI} (y_1)
= \int_{y_1}^{y_2}
dy \; J^y (+ \Delta x,y) -
J^y (- \Delta x,y)$.
The free energy per 2a-period,  normalized to the film thickness
and to $J_c \xi _0$
is given by the sum of three contributions: $F = f_{current} +
f_{field} + f_{Jos}$
where
$f_{current} = \int dxdy
[ \vec{J} (x,y) ]^2$, $f_{field} = \frac{\lambda^2}{d}
\int d^3 r \; \vec{H} ( \vec{r} )^2$, and
$f_{Jos}=
\bar{I} \int_{-a}^a
\frac{dy}{a}
g(y) \cos
[ \Delta \phi_{GI} (\theta) ]$
I have found that $f_{field} \leq f_{current}$,
so I neglect $f_{field}$ henceforth.
Using eqs. 8 and the Josephson boundary condition one sees that the free
energy has the same form as that previously
considered in eq. 5, except that the supercurrent term becomes
$$
f_{curr}= \frac{\pi\lambda}{4a}
\sum_p p^2
\frac{1+2b_p^2}{(1+b_p^2)^{3/2}}
[c_p^2 + d_p^2 ]
\eqno (9)
$$
instead of
$\frac{1}{2} \sum_n n (c_n^2 + d_n^2)$.

Note that for $b_p^2 \gg 1$, eq. 9 reduces to the previous
case, but for $b_p^2 \ll 1$, the supercurrent energy is much less
than found previously.
The previous analysis may now be applied with only minor modifications.
The weak coupling limit previously meant
$\bar{I} < 1$; now the condition is
$\bar{I} < \lambda /a$, which is much more stringent in high
$T_c$ materials where $\lambda \sim 1400 \AA$ and it is difficult
to imagine deliberately made devices with scales a $< 1 \mu m \sim 10^4 \AA$,
although as discussed below wandering of the A-B interface may
introduce a small length scale.
In this weak coupling limit the maximum current is
$\bar{I}^2 (a/ \lambda)$, and the solution corresponds to
vortices of extent $\lambda$ in the x direction and $a$ in the $y$
direction. In the range $\lambda /a < \bar{I} < (a/ \lambda )$ a linearized
equation cannot be used.  Proceeding  variationally one finds
that modes with
$p < p_{max} = \left( \frac{a \bar{I}}{\lambda} \right)^{1/2}$
have non-negligible amplitudes.
(Note that $b_{p\;max} < 1$ if
$\bar{I} < a/ \lambda$).
This implies that the scale of a vortex in the transverse
(y) direction is $\sim a/p_{max} \sim \left( \frac{\lambda a}{\bar{I}}
\right)^{1/2} \sim \left(\frac{\lambda \xi_0 J_c} {I_J} \right)^{1/2}$,
i.e. is the Josephson penetration depth of the junction.
The critical current is of order $I_J$
Finally, for $\bar{I} > a/\lambda$ the size of the vortex
becomes less than the Josephson penetration depth and the
problem becomes identical to the one previously considered;
the solution corresponds to vortices of scale $\lambda$ in the
transverse direction, pinched
where they cross the A-B interface.

I now turn to the experiment reported in ref. 1.
For $YBa_2Cu_3O_7$,
$\xi_0 \cong 10 \AA$ [3] and $I_J$
was measured [1] to be about
$2 \times 10^3 A/cm^2$, and the commonly accepted value is
$J_c = 2 \times 10^7 A/cm^2$.  Because the films were apparently
several thousand angrstoms thick, it seems reasonable to use the bulk value
$\lambda \sim 2 \times 10^{-5}$ cm for the penetration depth.
The circumference, $2 \pi a$, of the device used in ref. 1,
was 0.3 cm.
Using these numbers I find $\bar{I} \sim 10^2$ and $a/\lambda \sim 5 \time
10^3$.
It thus seems that the Josephson energy is dominant, but that the vortex size
along the interface is controlled by the Josephson penetration depth
$\lambda_J \sim 10^{-4} cm$.
Now in ref [1] "region A" was a hexagon and the orientations
were such that if the faces of the hexagon were
flat on an atomic scale, then the Josephson coupling for two of
the faces would vanish.  The considerations presented here would then
imply that would carry no supercurrent; the other four would carry a large
current.
In addition, the vortices at the corners of the hexagons should produce
a magnetic field which extends a distance $\lambda$ away from the corner
and $\lambda_J$ along the edge, and which at the surface would be of order
$\Phi_0/4\lambda \lambda_J \sim 10 gauss$.
In fact, the critical current through the individual faces was
measured by destroying one face after another (via laser ablation)
and monitoring the total critical current.
The face-to-face variation of the critical
current was found to be small, and for no face was the critical current found
to be zero.
This is not compatible with d-wave superconductivity if the interface is flat.
However, if  the interface wanders then the situation is different.  In
the rough interface case one might imagine that the Josephson coupling varies
as one moves along the interface.  If the length scale, L, over which the
coupling varies is larger than the Josephson penetration depth then the
previous arguments go though with the scale a replaced by L, and the
experimental results might be consistent with d-wave superconductivity.

Several extensions of the present work would be of use.
A detailed examination of the hexagonal geometry used in ref. 1 should
be performed.
Also, a numerical solution in the region
$\bar{I} \sim \bar{I}_c \cong 1.53$ would be of interest.
Finally, the dynamics of the system might be worth examining.
If the system is forced to have a current greater than the critical
current, voltage drops must occur.
Presumably the bulk superconducting regions will be equipotentials, so
there will be an angle-independent voltage drop across the A-B interface
and the vortices shown in fig. 2 might for small voltages precess about
the circle.
\section*{Acknowledgements}

I thank B. Batlogg, M. Beasley, J. Graybeal,  P. A. Lee, M. Sigrist
and C. M. Varma for helpful
conversations, L. Ioffe for drawing my attention to the importance of
the $\pi$-periodicity
found for small devices, M. Sigrist for suggestions on solving the
problem in the finite penetration
depth case, and E. Hellman for helpful discussions and a critical
reading
of the manuscript.
\newpage
\section*{References}
\bigskip
\begin{enumerate}
\item[1]
P. Chaudari and Y. Li, unpublished.
\item[2]
D. A. Wollman, D. J. van Haerlingen, W. C. Lee, D. M. Ginsberg and A. J.
Leggett,
Phys. Rev. Lett. {\it 71} 2134 (1993).
\item[3]
A. J. Millis, D. Rainer and J. Sauls,
Phys. Rev. B{\it 38} 4504 (1988).
\item[4]
M. Sigrist and T. M. Rice, J. Phys. Soc. Jpn. {\it 61} 4283 (1992).
\end{enumerate}
\newpage
\section*{Figure Captions}
\bigskip
\begin{itemize}
\item[Fig. 1]
System modelled in present paper shown for d-wave
to d-wave case.
Region A, inside the circle, is a film
of $d_{x^2-y^2}$ superconductor with crystal
axes running vertically and horizontally as shown
Region B, outside the circle, is a film of $d{_x^2-y^2}$
superconductor with crystal axes rotated by $45^{\circ}$.
The dots indicate places where the Josephson coupling across the circle
vanishes by symmetry; the plus and minus symbols indicate
the sign of the Josephson coupling in the regions between the dots.
In the d-s case the Josephson coupling would vanish only at
$0, \pi$ and $\pm \pi /2$.
\item[Fig. 2]
Pattern of induced currents calculated at
$\phi_0 = \pi / 2$ from eq. 9 for weak
Josephson coupling for s-wave to d-wave case.
Dots mark places where Josephson coupling vanishes.
Inset shows motion of centers of vortices in presence of
imposed outward current.
\item[Fig. 3]
Linear geometry used to study finite $\lambda$ case.
Dots mark points at which a Josephson coupling changes sign.
The $+/-$ labels the local sign of Josephson coupling across interface.
The dotted line is the contour used in the computation
of $\Delta \phi_{GI}$.
\end{itemize}
\end{document}